\documentclass[aps,preprint,prl,superscriptaddress]{revtex4}

\usepackage{graphicx}
\usepackage{amsmath}
\usepackage{amssymb}
\usepackage{units}
\usepackage{color}
\usepackage{float}

\usepackage{setspace}

\begin{document}

\title{Band Structure Dynamics in Indium Wires}

\author{M. Ch{\'a}vez-Cervantes}
\email{mariana.chavez-cervantes@mpsd.mpg.de}
\author{R. Krause}
\author{S. Aeschlimann}
\author{I. Gierz}
\email{isabella.gierz@mpsd.mpg.de}
\affiliation{Max Planck Institute for the Structure and Dynamics of Matter, Center for Free Electron Laser Science, Hamburg, Germany}

\date{\today}

\begin{abstract}
One-dimensional Indium wires grown on Si(111) substrates, which are metallic at high temperatures, become insulating below $\sim100$\,K due to the formation of a Charge Density Wave (CDW). The physics of this transition is not conventional and involves a multiband Peierls instability with strong interband coupling. This CDW ground state is readily destroyed with femtosecond laser pulses resulting in a light-induced insulator-to-metal phase transition. The current understanding of this transition remains incomplete, requiring measurements of the transient electronic structure to complement previous investigations of the lattice dynamics. Time- and angle-resolved photo\-emission spectroscopy with extreme ultra-violet radiation is applied to this end. We find that the transition from the insulating to the metallic band structure occurs within $\sim660$\,fs that is a fraction of the amplitude mode period. The long life time of the transient state ($>100$\,ps) is attributed to trapping in a metastable state in accordance with previous work.
\end{abstract}

\maketitle

Charge density wave (CDW) order \cite{GrünerBook1994} arises in low-dimensional solids due to the presence of large parallel sections on the Fermi surface separated by a nesting vector q$_{\text{CDW}}$. A periodic rearrangement of the charges with a real space periodicity of 2$\pi$/q$_{\text{CDW}}$ opens up a band gap at the Fermi level resulting in a metal-to-insulator phase transition. This transition is driven by strong electron-phonon coupling (Peierls transition) and/or strong electronic correlations (Mott transition). It is typically discussed in terms of a complex order parameter, the amplitude of which represents the gap in the spectral function. Notably, CDW order (as well as other symmetry-broken ground states of solids) is readily destroyed with light, as quasiparticles are excited across this gap \cite{CavalleriPhysRevLett2001, IwaiPhysRevLett2003, CavalleriPhysRevB2004, CholletScience2005, TomeljakPhysRevLett2009, EichbergerNature2010, HellmannPhysRevLett2010}. The observed time scales for CDW melting reveal important information about the microscopic physics of the CDW transition \cite{SohrtFaradayDiscuss2014}. Time- and angle-resolved photoemission spectroscopy (tr-ARPES) allows for a momentum-, energy-, and time-resolved investigation of these phenomena, and highlights amplitude and phase oscillations of the order parameter, observed as periodic modulations of the size and the position of the band gap in momentum space, respectively \cite{PerfettiPhysRevLett2006, SchmittScience2008, RohwerNature2011, PetersenPhysRevLett2011, HellmannNatCommun2012, LiuPhysRevB2013, MathiasNatCommun2015, RettigNatCommun2015, GrünerBook1994}.

Here we use tr-ARPES to investigate the light-induced insulator-to-metal transition in one-dimensional Indium wires grown {\it in situ} on a Si(111) substrate. The structural dynamics of this transition have been revealed in previous time-resolved reflection high-energy electron diffraction (tr-RHEED) experiments \cite{WallPhysRevLett2012, FriggeNature2017}. Complementary measurements of the electronic structural dynamics are required for a complete picture of the light-induced phase transition. 

\begin{figure}
	\center
  \includegraphics[width = 0.5\columnwidth]{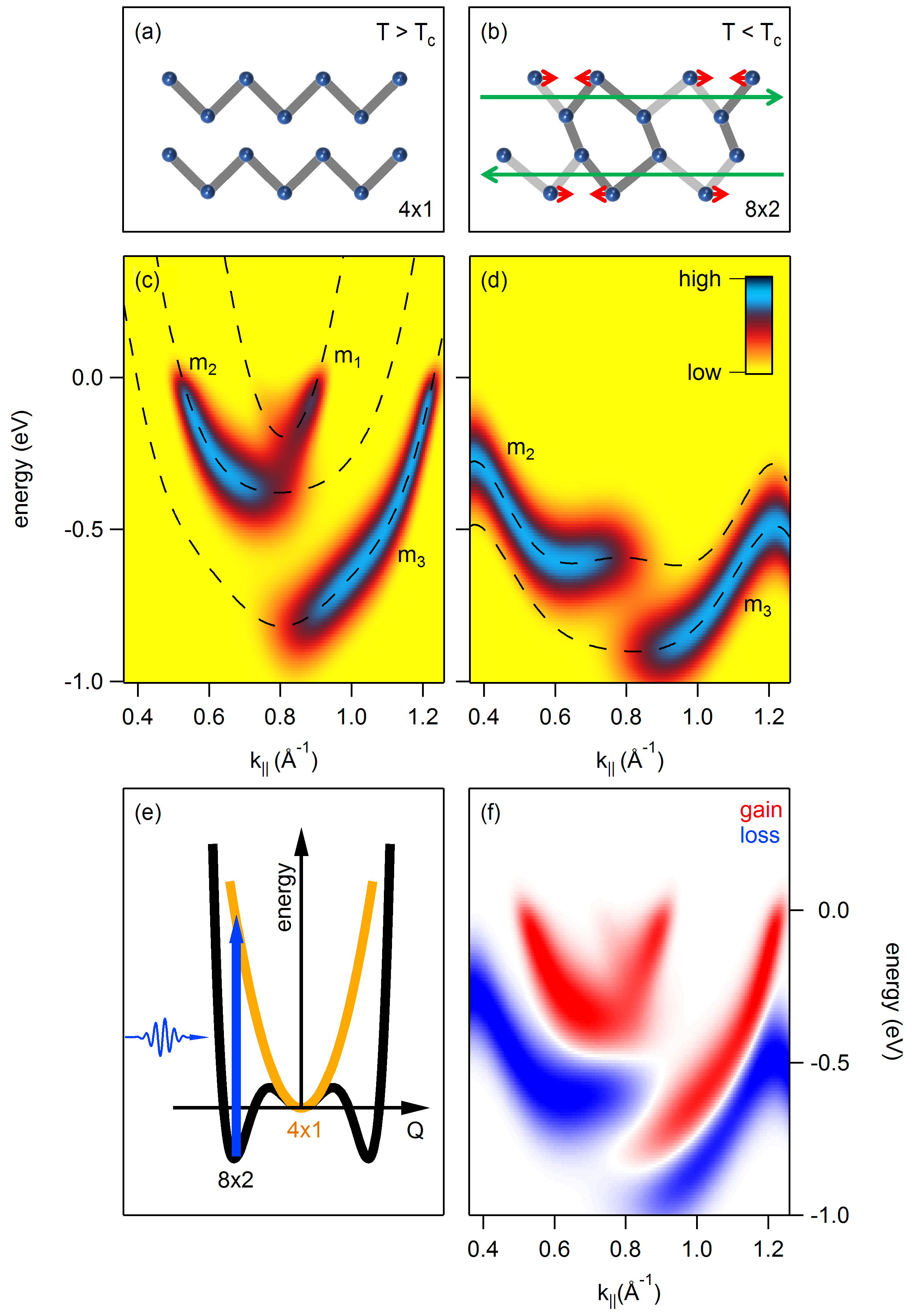}
  \caption{Metal-to-insulator transition in 1D Indium wires on Si(111): Sketch of the atomic structure above (a) and below T$_{\text{C}}$ (b). Sketch of the electronic structure above (c) and below T$_{\text{C}}$ (d) based on density functional theory calculations from \cite{KimPhysRevB2016}. (e) Sketch of the potential energy surface below (black) and above T$_{\text{C}}$ (orange). Q is the generalized reaction coordinate of the ($4\times1$) to ($8\times2$) transition as defined in \cite{WallPhysRevLett2012}. The vertical blue arrow indicates the light-induced phase transition. (f) Difference between panel (c) and (d) simulating the pump-probe signal for the light-induced phase transition.}
  \label{fig1}
\end{figure}

At room temperature (Fig. \ref{fig1}a) the Indium atoms form pairs of ``zigzag'' chains separated by one chain of Silicon atoms, resulting in a (4$\times$1) unit cell \cite{BunkPhysRevB1999}. This crystal structure gives rise to three metallic bands (Fig. \ref{fig1}c) that cross the Fermi level at k$_{\text{F}}$= 0.75${\AA}^{-1}$ (m$_1$), k$_{\text{F}}$= 0.54${\AA}^{-1}$ (m$_2$), and k$_{\text{F}}$= 0.41${\AA}^{-1}$ (m$_3$) corresponding to band fillings of 0.11 (m$_1$), 0.38 (m$_2$), and 0.50 (m$_3$) \cite{AbukawaSurfSci1995, AhnPhysRevLett2004}. The sketches in Figs. \ref{fig1}c and d are based on density functional theory calculations from \cite{KimPhysRevB2016} and include photo\-emission matrix element effects that suppress the right branch of m$_2$ and the left branch of m$_1$ and m$_3$, respectively \cite{MorikawaPhysRevB2010}.

Below a critical temperature of about 100\,K (Fig. \ref{fig1}b) a periodic lattice distortion drives the structure into a (8$\times$2) phase, accompanied by a metal-to-insulator transition \cite{YeomPhysRevLett1999, AhnPhysRevLett2004, GonzalezNJPhys2004, SunPhysRevB2008}. The dimerization of the outer Indium atoms (red arrows in Fig. \ref{fig1}b) results in a doubling of the unit cell along the wires which gaps the half-filled band m$_3$. In addition, a shear distortion displaces neighboring Indium chains along the direction parallel to the chains (green arrows in Fig. \ref{fig1}b) which transfers charge from m$_1$ to m$_2$. As a result, m$_1$ becomes unoccupied and m$_2$ almost half-filled. Due to the dimerization, m$_2$ also gaps. The size of the band gap below the Fermi level is 0.34\,eV for m$_3$ and ranges between 0.04 and 0.09\,eV depending on the momentum for m$_2$ \cite{AhnPhysRevLett2004, SunPhysRevB2008}. The semiconducting band structure is sketched in Fig. \ref{fig1}d. The system possesses two amplitude modes, a shear mode at 0.6\,THz and a rotary mode which includes the dimerization of the outer Indium atoms at 0.8\,THz, that soften when approaching the critical temperature \cite{JeckelmannPhysRevB2016}. 

In Fig. \ref{fig1}e we show the potential energy surfaces for the (4$\times$1) and (8$\times$2) structures in orange and black, respectively, where Q is the generalized reaction coordinate of the structural phase transition as defined in \cite{WallPhysRevLett2012}. The potential energy surface for the (8$\times$2) phase deviates from the standard mexican hat shape and exhibits an additional minimum at Q = 0 that corresponds to a metastable (4$\times$1) phase that is separated from the true (8$\times$2) ground state by a potential barrier of 40\,meV \cite{WallPhysRevLett2012,JeckelmannPhysRevB2016}. Photoexcitation across the CDW gap is expected to bring the system from the black to the orange potential energy surface along the blue arrow. The expected tr-ARPES pump-probe signal for this transition is sketched in Fig. \ref{fig1}f. This sketch is obtained by subtracting the T $<$ T$_{\text{C}}$ band structure (Fig. \ref{fig1}c) from the T $>$ T$_{\text{C}}$ band structure (Fig. \ref{fig1}d).

Details about the sample preparation and the tr-ARPES setup are given in the Supplemental Material \cite{SupMat}.

Figure \ref{fig2} shows the equilibrium band structure at T = 40\,K (Fig. \ref{fig2}a) together with the response of the band structure to photoexcitation of the system with femtosecond pulses at 1.0\,eV photon energy with a fluence of 5.6\,mJ/cm$^2$ (Fig. \ref{fig2}b). At the peak of the electronic excitation (t = 0.1\,ps) we observe a broad distribution of photoexcited electrons and holes. After 1.1\,ps, the pump-probe signal is mainly confined to the area below the Fermi level and closely resembles the sketch in Fig. \ref{fig1}f for a light-induced insulator-to-metal phase transition. We observe no further relaxation within 3\,ps, indicating a long-lived transient state.

In order to quantify the CDW melting and relaxation dynamics we integrate the pump-probe signal over the areas indicated by the black boxes in Fig. \ref{fig2}b yielding the pump-probe traces in Fig. \ref{fig3}a. From exponential fits to the data (thick lines in Fig. \ref{fig3}a, for details see \cite{SupMat}) we determine the rise and decay times shown in Fig. \ref{fig3}b and c, respectively. We attribute the short-lived pump-probe signal at high energies to the electronic excitation. The long-lived response ($>100$\,ps) with the slow rise time ($660\pm180$\,fs) below the Fermi level reflects the melting of the CDW: The band gaps in m$_2$ and m$_3$ close and m$_1$ shifts back below the Fermi level.

For comparison, we present tr-ARPES data taken at room temperature in the (4$\times$1) phase in Fig. \ref{fig4}. Figures \ref{fig4}a, b, and c show the equilibrium photocurrent, the simulated pump-probe signal assuming a hot electronic distribution for the transient state, and the measured pump-probe signal for pump-probe time delays of 0.1 and 1.0\,ps for a pump photon energy of 1.55\,eV and a pump fluence of 2\,mJ/cm$^2$, respectively. We confirmed that the difference in pump photon energy and fluence with respect to the data taken at 40\,K does not affect the conclusions drawn in this section \cite{SupMat}. In contrast to the data for the (8$\times$2) phase in Figs. \ref{fig2} and \ref{fig3}, the pump-probe signal for the (4$\times$1) phase is almost gone after 1\,ps. Individual pump-probe traces taken at different energies (Fig. \ref{fig4}d) show only a small increase of the rise (decay) time from $130\pm5$\,fs ($590\pm10$\,fs) to $240\pm20$\,fs ($1.10\pm0.03$\,ps) when approaching the Fermi level (see red data points in Fig. \ref{fig3}b and c). The close resemblance between Figs. \ref{fig4}b and c indicates that the pump-probe signal can be understood in terms of a hot electronic distribution characterized by the time-dependent electronic temperature shown in Fig. \ref{fig4}e (for details see \cite{SupMat}) without any underlying changes of the band structure. The cooling time for the hot electrons is found to be $1.18\pm0.05$\,ps from exponential fitting. 

Next, we compare our results to previous tr-RHEED measurements of the lattice dynamics \cite{FriggeNature2017,WallPhysRevLett2012}. In Ref. \cite{FriggeNature2017}, the (8$\times$2) to (4$\times$1) transition was found to occur within 350\,fs (680\,fs according to our definition of the rise time, see \cite{SupMat}) in excellent agreement with our own value of $660\pm120$\,fs. This number is slightly longer than 1/4 period of the two amplitude modes (310\,fs and 420\,fs, respectively \cite{JeckelmannPhysRevB2016}) which set the lower limit for CDW melting in a Peierls picture. The recovery time of the (8$\times$2) phase was found to be 440\,ps with tr-RHEED \cite{WallPhysRevLett2012}, consistent with our value of $>$100\,ps for the recovery of the semiconducting band structure. 

The long life time of the transient (4$\times$1) phase in the tr-RHEED experiments was attributed to trapping in the metastable (4$\times$1) phase shown in Fig. \ref{fig1}e \cite{WallPhysRevLett2012}. Relaxation into the (8$\times$2) phase was proposed to happen via condensation nuclei that grow with a velocity of 850\,m/s \cite{WallPhysRevLett2012}.

In agreement with Ref. \cite{FriggeNature2017} we find no indications for coherent amplitude mode oscillations in our data below T$_{\text{C}}$. This has been attributed to strong damping of the amplitude modes due to rapid energy transfer to various phonon modes of the surface and the substrate \cite{FriggeNature2017}. However, we observe coherent oscillations in the pump-probe trace centered at $+0.1$\,eV at room temperature in Fig. \ref{fig4}d. The frequency of this oscillation, determined from a sinusoidal fit, is $1.8\pm0.1$\,THz. Previous Raman measurements show a peak at the same frequency that was attributed to a phonon that periodically modulates the distance between two neighboring Indium chains \cite{FleischerPhysRevB2007, SpeiserPhysRevB2016}.

In summary, we have shown that the transition from the insulating to the metallic band structure in one-dimensional Indium wires on Si(111) occurs on a time scale slightly longer than 1/4 period of the amplitude modes of the system consistent with a Peierls transition. After photoexcitation the system gets trapped in a metastable ($4\times1$) phase with $>100$\,ps life time. Our findings are in excellent agreement with previous complementary investigations of the lattice dynamics with tr-RHEED \cite{WallPhysRevLett2012,FriggeNature2017}.

We thank H. Bromberger for technical support and A. Cavalleri for careful reading of the manuscript and many valuable comments. This work received financial support from the German Science Foundation via the SFB 925 ``Light induced dynamics and control of correlated quantum systems''.

\clearpage
\pagebreak

\begin{figure}
	\center
  \includegraphics[width = 1\columnwidth]{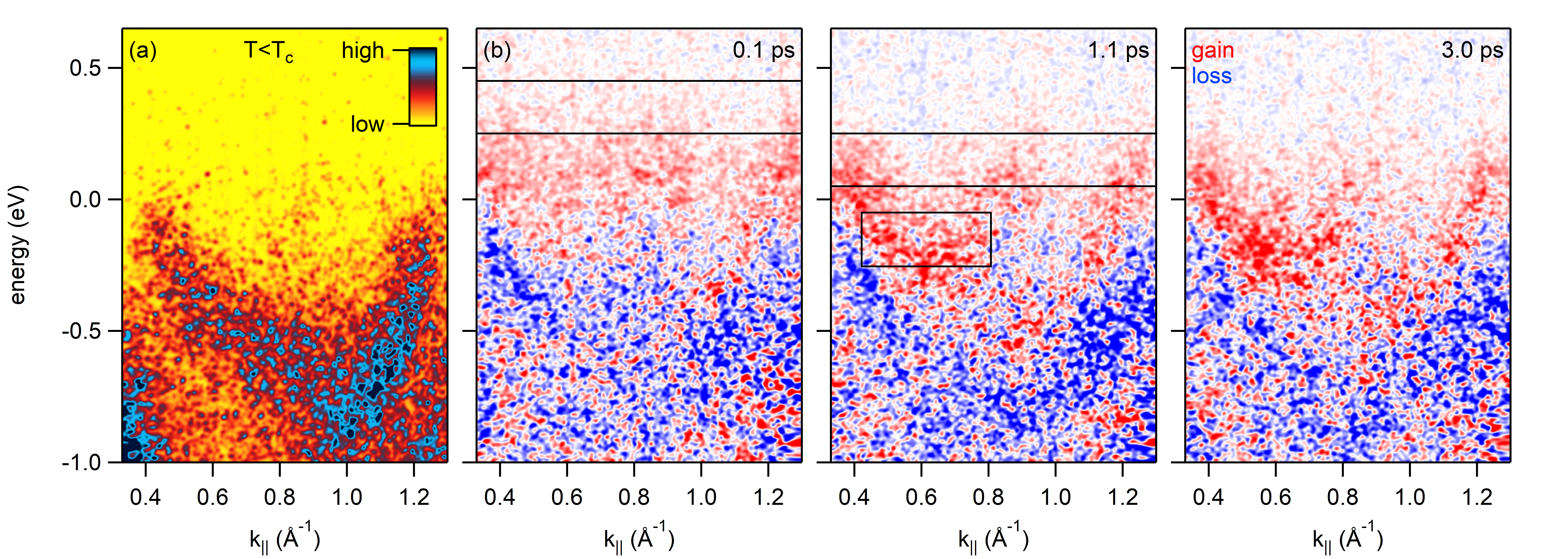}
  \caption{Light-induced insulator-to-metal transition viewed by tr-ARPES at T = 40\,K. (a) Equilibrium photocurrent. (b) Pump-induced changes of the photocurrent at t = 0.1\,ps, t = 1.1\,ps, and t = 3.0\,ps for a pump photon energy of 1.0\,eV and a pump fluence of 5.6\,mJ/cm$^2$. The black boxes (centered at $\pm0.15$\,eV and $+0.35$\,eV) indicate the area of integration for the pump-probe traces in Fig. \ref{fig3}a.}
  \label{fig2}
\end{figure}

\begin{figure}
	\center
  \includegraphics[width = 0.75\columnwidth]{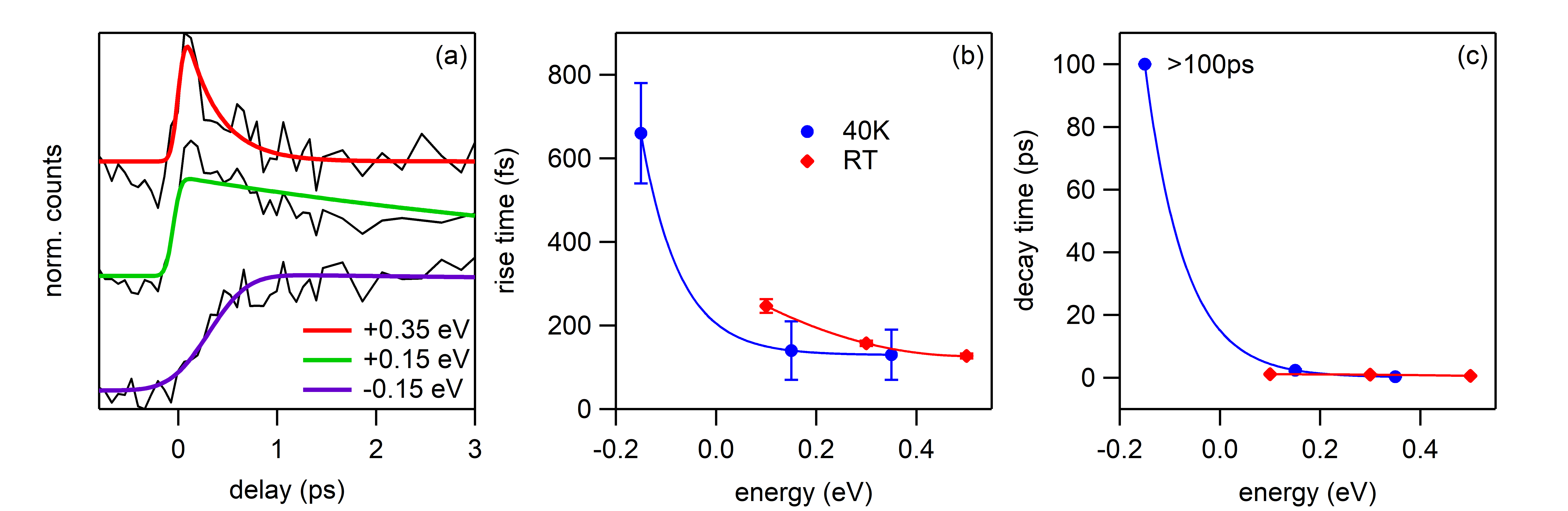}
  \caption{Energy dependence of the pump-probe signal at T = 40\,K. (a) Pump-probe traces (thin lines) obtained by integrating the pump-induced changes of the photocurrent over the areas indicated by black boxes in Fig. \ref{fig2}b. Thick lines are exponential fits. (b) Rise and (c) decay time as a function of energy. The values for the room temperature (RT) measurement from Fig. \ref{fig4} are included in (b) and (c) for direct comparison. Continuous lines in (b) and (c) are guides to the eye.}
  \label{fig3}
\end{figure}

\begin{figure}
	\center
  \includegraphics[width = 0.5\columnwidth]{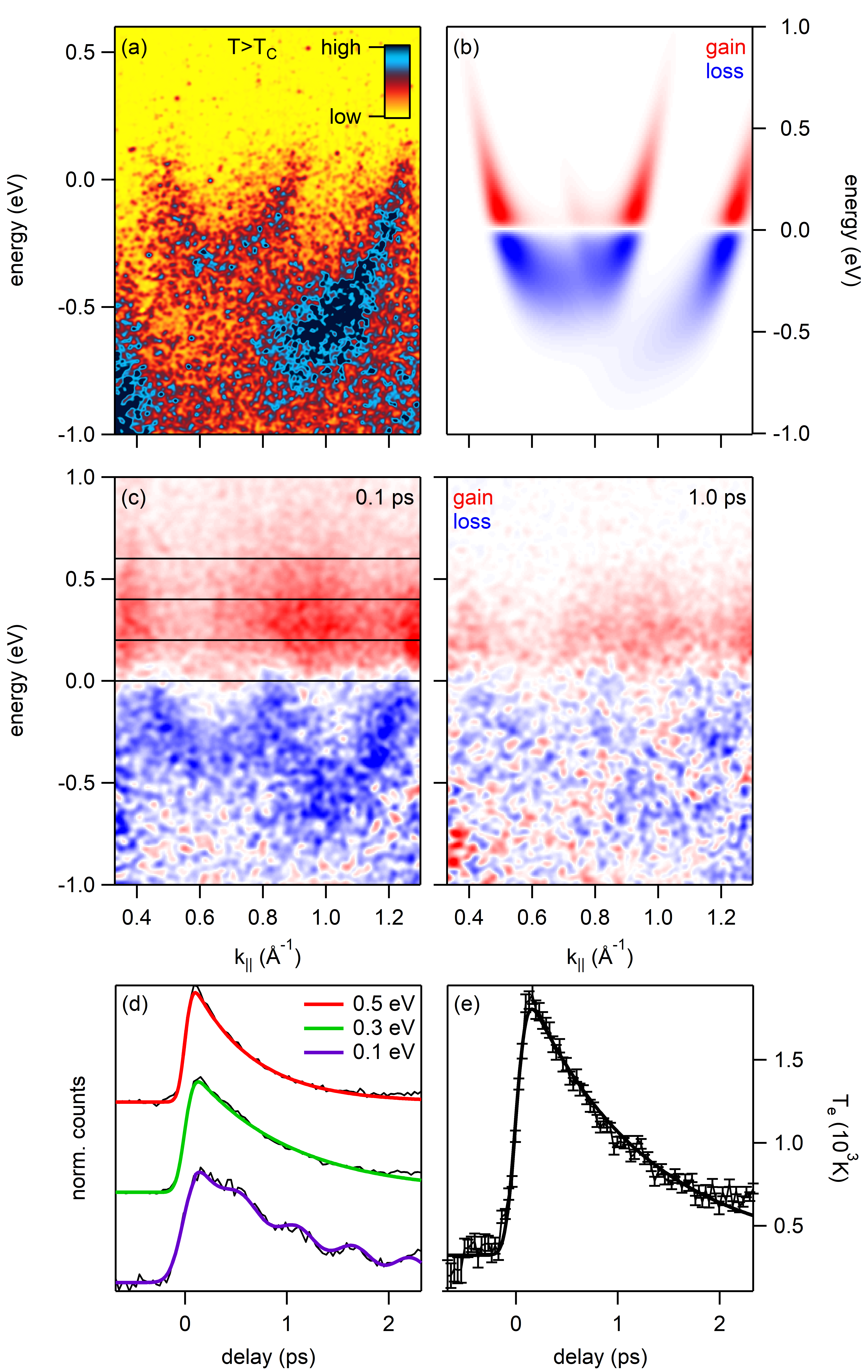}
  \caption{Dynamics at room temperature. (a) Equilibrium photocurrent. (b) Simulated pump-probe signal assuming a hot electronic distribution for the transient state. (c) Pump-induced changes of the photocurrent at t = 0.1\,ps and t = 1.1\,ps for a pump photon energy of 1.55\,eV and a pump fluence of 2\,mJ/cm$^2$. The black boxes indicate the area of integration for the pump-probe traces in panel (d). Thick lines in (d) are exponential fits. The pump-probe trace at 0.1\,eV shows coherent oscillations at $1.8\pm0.1$\,THz. (e) Electronic temperature as a function of pump-probe delay together with exponential fit.}
  \label{fig4}
\end{figure}

\clearpage
\pagebreak

\noindent {\bf Supplemental Material}\\

\noindent {\bf Sample Preparation}\\

\begin{figure}[H]
	\center
  \includegraphics[width = 1\columnwidth]{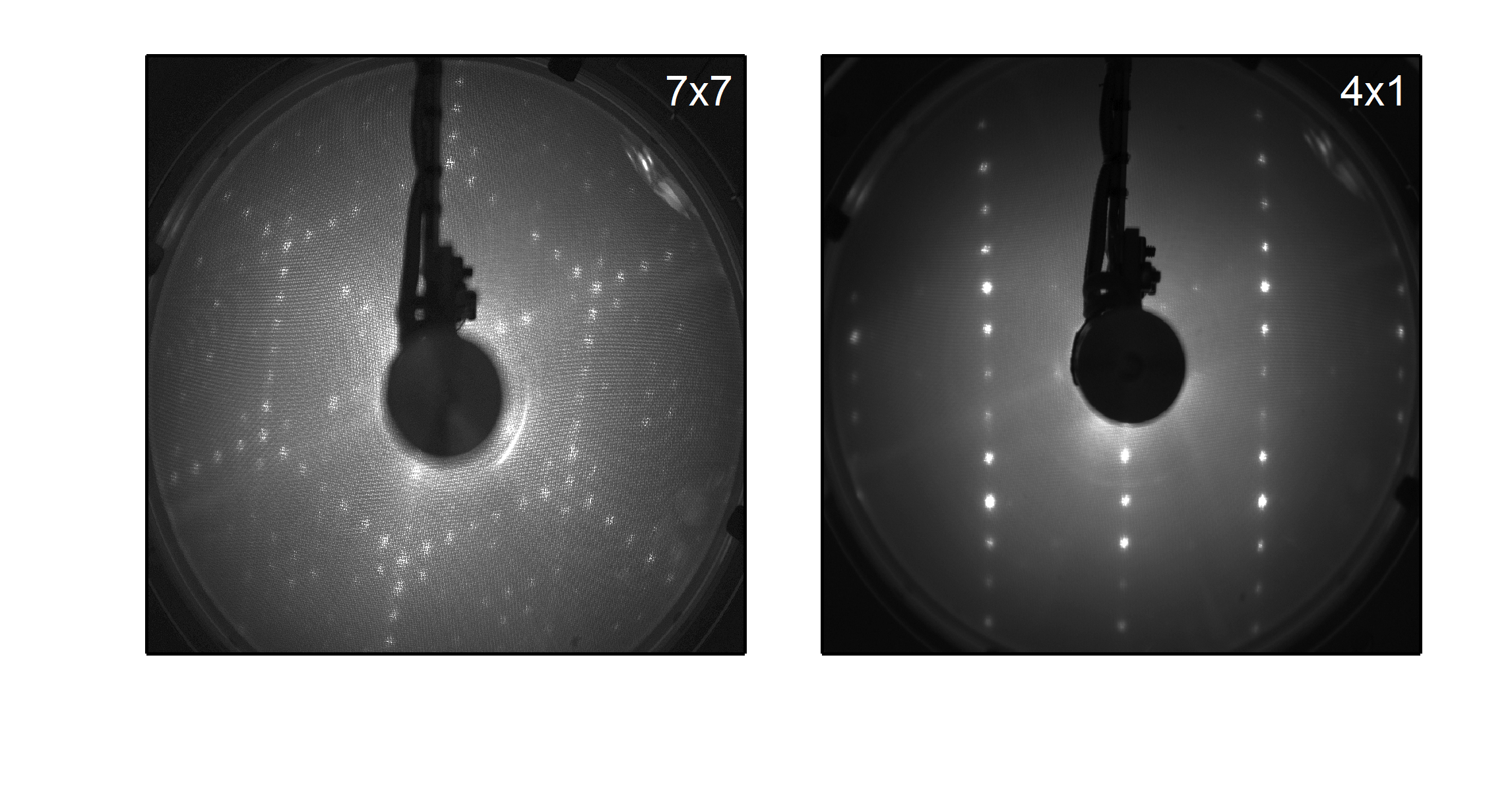}
  \caption{LEED pictures of the Si(111) ($7\times7$) and the Indium covered ($4\times1$) surfaces taken at 58\,eV and 77\,eV, respectively.}
  \label{fig5}
\end{figure}

\noindent The samples were prepared following the procedure detailed in \cite{StevensPhysRevB1993, AbukawaSurfSci1995, SakamotoPhysRevB2000}. A Si(111) substrate with a small miscut was used to ensure the growth of Indium wires in a single (4$\times$1) domain. Phosphorous-doped Si(111) with a resistance $<0.01$\,$\Omega$cm and a 1.5$^{\circ}$ miscut along the [$\overline{1}\overline{1}2$] direction was annealed to 1100\,$^{\circ}$C for 10 minutes. Afterwards the temperature was slowly reduced to 850\,$^{\circ}$C. These steps were repeated until the pressure stayed below 10$^{-9}$\,mbar. We then flashed the substrate to 1260\,$^{\circ}$C and slowly cooled down to 1060\,$^{\circ}$C followed by a fast temperature decrease to 850\,$^{\circ}$C to obtain regular steps of monoatomic height on the ($7\times7$) surface \cite{LinJApplPhys1998}. Next we deposited $\sim10$ monolayers of Indium on the substrate at room temperature from an electron beam evaporator and annealed the sample at 500\,$^{\circ}$C for 5 minutes resulting in the desired ($4\times1$) structure. All steps during sample preparation were monitored with low energy electron diffraction (LEED, Fig. \ref{fig5}).

\noindent {\bf Tr-ARPES Setup}\\

\noindent The tr-ARPES setup was based on a Titanium:Sapphire amplifier operating at a repetition rate of 1\,kHz. 2\,mJ of output power were used for high harmonics generation in Argon, producing a broad spectrum of extreme ultra-violet (XUV) light ranging from 10 to 45\,eV photon energy. A single harmonic at $\hbar\omega_{\text{probe}}=26$\,eV was selected with a time-preserving grating monochromator \cite{FrassettoOptExpress2011} and used as a probe pulse for the tr-ARPES experiments. This photon energy was high enough to reach beyond the first Brillouin zone boundary of the In/Si(111) ($4\times1$) phase and measure the complete band structure of the system. After excitation of the sample with a pump pulse ($\hbar\omega_{\text{pump}}=1.0$\,eV or 1.55\,eV) the time-delayed XUV pulse ejected photoelectrons that passed through a hemispherical analyzer after which snapshots of the one-dimensional band structure were obtained on a two-dimensional detector. The overall energy and time resolution of the tr-ARPES experiment were 300\,meV and 80\,fs, respectively.\\

\noindent {\bf Data Analysis}\\

\noindent We used a step edge plus an exponential decay convolved with a gaussian to fit the pump-probe traces in Fig. 3a and Fig. 4d and the electronic temperature in Fig. 4e of the manuscript. The fitting function is given by 

$$I(t)=c+\frac{a}{2} \left[1+erf\left(\frac{(t-t_0)/\tau-(rt/2.355)^2}{\sqrt{2} (rt/2.355)\tau}\right)\right]exp\left(\frac{(rt/2.355)^2-2(t-t_0)\tau}{2\tau^2}\right),$$

\noindent where $c$ is a constant offset, $a$ is the amplitude, $erf$ is the error function, $t_0$ is the  middle of the rising edge, and $\tau$ is the decay time. The rise time $rt$ is defined as the full width at half maximum of the temporal derivative of the rising edge.

\begin{figure}[H]
	\center
  \includegraphics[width = 0.5\columnwidth]{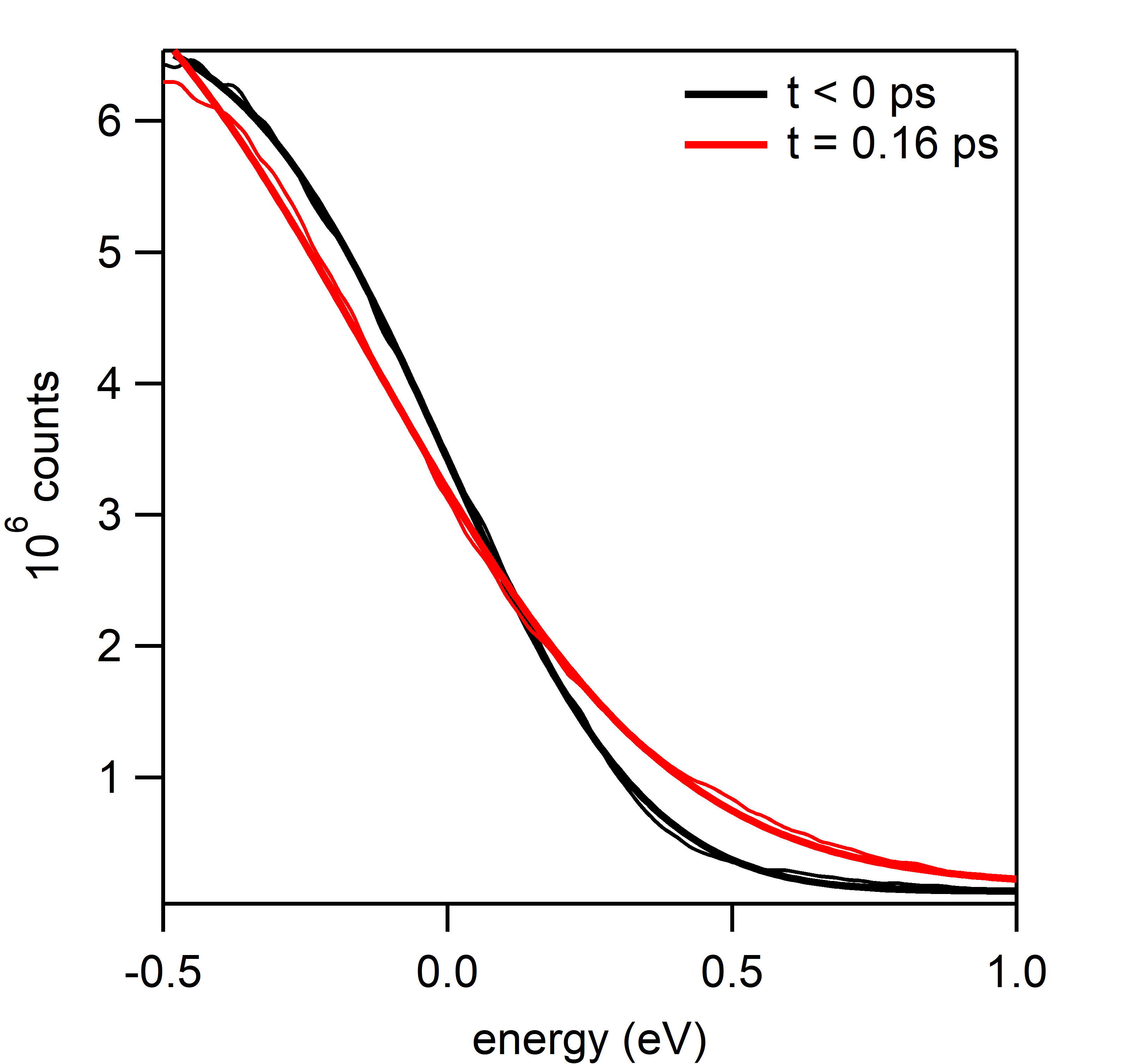}
  \caption{EDCs (thin lines) at negative delay and at t = 160\,fs together with Fermi-Dirac fits (thick lines).}
  \label{fig6}
\end{figure}

\noindent The electronic temperature in Fig. 4e of the manuscript was determined as follows: First, the photocurrent was integrated over the whole momentum axis shown in Fig. 4a and c, yielding a single energy distribution curve (EDC) for each pump-probe time delay. These EDCs were then fitted with a Fermi-Dirac distribution (see Fig. \ref{fig6}).

\noindent {\bf Room temperature tr-ARPES data for 1\,eV pump photon energy}\\

\begin{figure}[H]
	\center
  \includegraphics[width = 1\columnwidth]{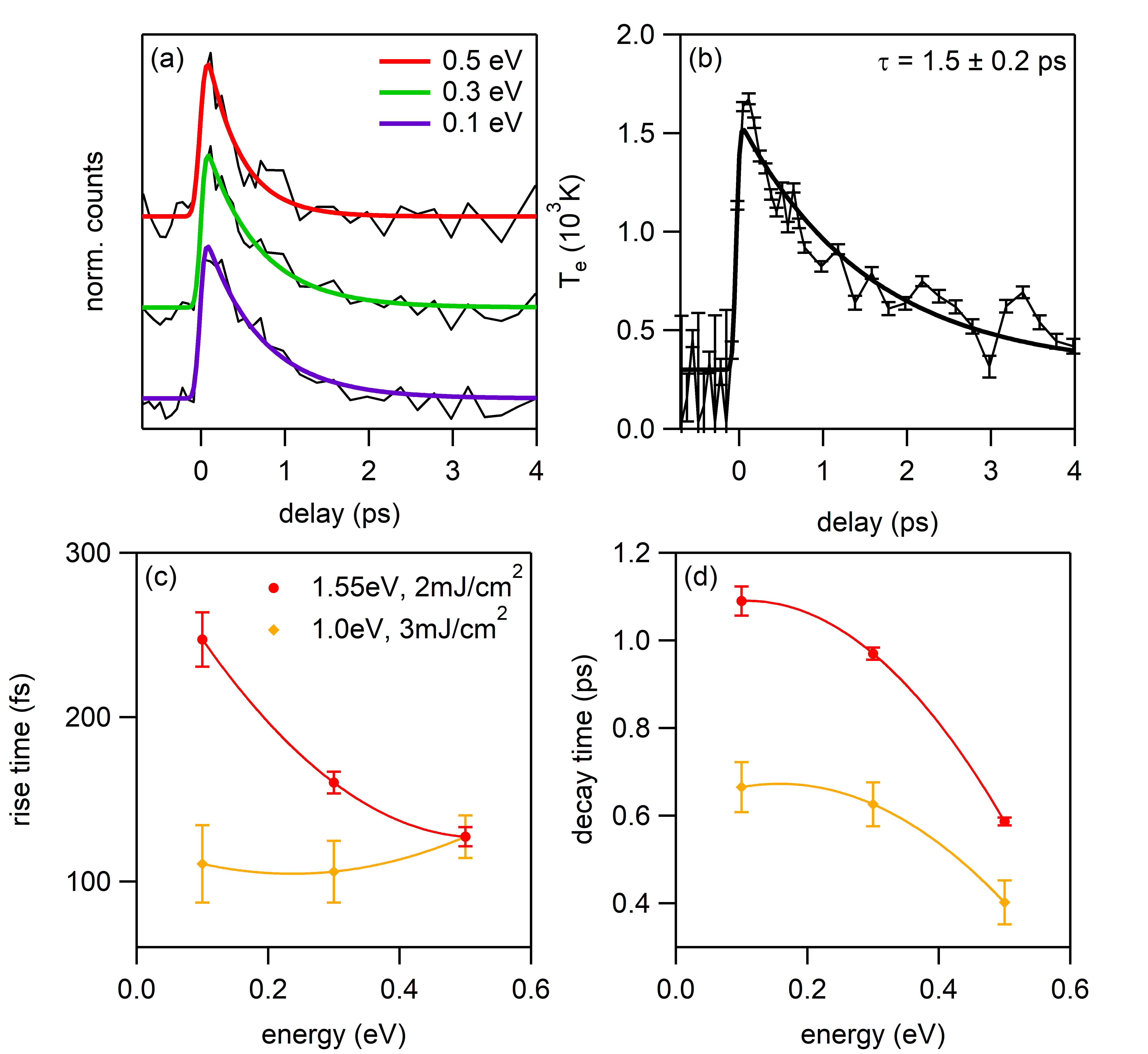}
  \caption{Dynamics at room temperature for 1\,eV pump pulses with a fluence of 3\,mJ/cm$^2$. (a) Pump-probe traces for different energies together with exponential fits. (b) Electronic temperature as a function of pump-probe delay together with exponential fit. (c) Rise and (d) decay time as a function of energy for photoexcitation at 1\,eV (orange) and 1.55\,eV (red) pump photon energy. Continuous lines in (c) and (d) are guides to the eye.}
  \label{fig7}
\end{figure}

\noindent To ensure that photoexcitation at 1.55\,eV gives similar results as photoexcitation at 1.0\,eV we repeated the room temperature measurements with a pump photon energy of 1.0\,eV. The results are presented in Fig. \ref{fig7} for direct comparison with Figs. 3 and 4 in the main text.

\end{document}